# An Interpretable Deep Learning Framework for Discovery and Clinical Validation of Deep Radiomic Signatures in Tumor Classification

Chengkun Sun[1], Jinqian Pan[1], Renjie Liang[1], Zhengkang Fan[1], Xin Miao[2], Yi Guo[1], Mei Liu[1], Muxuan Liang[3] , Russell Terry[4], and Jie Xu[1*]

[1*] Department of Health Outcomes and Biomedical Informatics, University of Florida, Gainesville, 32611, FL, USA.

[2]Tiktok, San Jose, 95110, CA, USA.

[3]Department of Biostatistics, MD Anderson Cancer Center, Houston, 77030, TX, USA.

[4]Department of Urology, University of Florida, Gainesville, 32611, FL, USA

*Corresponding author(s). E-mail(s): xujie@ufl.edu; Contributing authors: sun.chengkun@ufl.edu;

**Abstract**

Imaging signatures are quantitative features extracted from medical images that provide clinically meaningful information for tumor diagnosis, characterization, prognosis, and treatment planning. Although deep learning has shown great potential for imaging signature discovery, its limited interpretability remains a major barrier to clinical adoption. Existing approaches often achieve high predictive performance but provide little biological insight into the identified signatures. We propose a unified framework for interpretable imaging signature discovery by integrating deep learning–based segmentation, explainable classification, and radiomic analysis. A robust segmentation model is first used to accurately delineate tumors, followed by a Grad-CAM guided pipeline that identifies diagnostically important regions as candidate imaging signatures. A mutual information based adaptive thresholding strategy enables patient-specific signature extraction. The resulting signatures are validated using a downstream deep learning classification model, while radiomic features extracted from the signature regions are evaluated with traditional machine learning models and interpreted using SHAP to identify the most discriminative biomarkers. The proposed framework is evaluated on the public BUSI breast ultrasound, KiTS renal CT, and BraTS brain tumor datasets, as well as a private UF Health renal CT cohort. Compared with conventional whole-tumor radiomics, the proposed signature-based approach achieves improved discriminative performance while providing greater biological interpretability. By converting deep learning attention into reproducible quantitative imaging biomarkers, this framework offers an interpretable and reproducible solution for non-invasive tumor characterization and imaging biomarker discovery.

# 1 Introduction

Tumors represent a major and growing global health burden, with substantial morbidity and mortality across a wide range of cancer types [1]. According to the latest estimates from the World Health Organization and the International Agency for Research on Cancer, there were approximately 20 million new cancer cases and 9.7 million cancer-related deaths worldwide in 2022 alone, with projections exceeding 35 million new cases annually by 2050, representing a 77% increase [1]. Additionally, about 1 in 5 people worldwide will develop cancer during their lifetime, underscoring the escalating global burden of malignancy. Early and accurate characterization of tumors is essential for guiding clinical decision-making, including distinguishing benign from malignant lesions and identifying clinically relevant subtypes [2]. However, despite advances in imaging technologies such as computed tomography (CT) and magnetic resonance imaging (MRI), a large proportion of lesions remain indeterminate in routine clinical [3]. This challenge is particularly pronounced for small or heterogeneous tumors, where overlapping imaging appearances can obscure subtle but biologically meaningful distinctions and limit reliable diagnosis using conventional visual assessment alone [4]. As a result, tumor classification based solely on standard imaging often remains insufficient, frequently necessitating additional invasive procedures such as biopsy or surgical resection for definitive characterization [5]. In current clinical workflows, tumor characterization typically relies on radiological interpretation supported by pathological confirmation [6]. Meanwhile, radiological assessment is largely qualitative and subjective, leading to inter-observer variability across institutions and readers [7]. These limitations frequently result in diagnostic uncertainty, which can lead to overtreatment of benign lesions or delayed treatment of aggressive disease [8]. Therefore, there is an increasing need for interpretable computational approaches that not only enhance diagnostic accuracy but also uncover the underlying imaging biomarkers and biologically relevant regions that drive classification decisions, thereby improving clinical trust and enabling more transparent, noninvasive decision-making [9, 10].

Recent advances in deep learning have demonstrated substantial potential in medical image analysis, particularly for segmentation, classification, and detection tasks [11]. By learning hierarchical feature representations directly from large-scale imaging data, deep neural networks can capture complex anatomical and pathological patterns that often extend beyond conventional human visual assessment [12]. In tumor classification, these capabilities offer a promising pathway toward noninvasive, automated, and scalable clinical decision-support systems [13]. Traditionally, tumor characterization has relied heavily on radiomics-based signatures derived from handcrafted features extracted from manually or automatically segmented tumor regions [14]. These conventional signatures typically summarize global tumor characteristics such as shape, intensity, and texture across the entire lesion, providing interpretable but often coarse representations of tumor phenotype [15]. While such handcrafted signatures can offer biological transparency, they generally assume that the whole tumor contributes uniformly to diagnosis, potentially overlooking spatially localized yet clinically critical subregions such as infiltrative margins, necrotic cores, hypervascular niches, or heterogeneous transition zones [16]. Consequently, traditional radiomic signature construction often depends heavily on accurate tumor segmentation, handcrafted feature engineering, and extensive expert-driven preprocessing, making the overall workflow time-consuming, labor-intensive, and potentially difficult to scale efficiently in routine clinical practice [17]. In contrast, deep learning models can rapidly and automatically generate highly discriminative feature signatures directly from imaging data without the need for predefined feature engineering, substantially improving efficiency, scalability, and predictive performance [18]. By leveraging multiscale spatial relationships and complex nonlinear imaging patterns, these approaches often outperform traditional handcrafted pipelines while reducing the time and labor associated with manual feature design [19]. However, this efficiency and performance advantage comes with a major limitation: deep learning–derived signatures are often biologically opaque [20]. Because the learned representations are embedded within high-dimensional latent spaces, it is often difficult to determine which specific tumor regions, imaging biomarkers, or pathological characteristics drive classification decisions [21]. As a result, although deep learning enables fast and powerful predictive modeling, its signatures frequently lack clear interpretability and clinical transparency, limiting trust and hindering reliable translation into practice [22]. This creates a fundamental gap between traditional and deep learning–based signature paradigms [22]. Traditional signatures are generally interpretable but may inadequately capture localized biological complexity, whereas deep learning signatures are often highly predictive but insufficiently explainable [22]. Bridging this gap is particularly important in tumor classification, where clinically meaningful imaging patterns often arise from biologically distinct subregions rather than from homogeneous whole-tumor representations [23]. Therefore, there is a growing need for computational frameworks that can preserve the predictive strength of deep learning while explicitly discovering interpretable imaging signatures linked to biologically relevant tumor subregions [24].

Currently, interpretability methods for deep learning in medical imaging primarily encompass a variety of strategies aimed at revealing how models make predictions and the underlying decision-making mechanisms [25]. Concept learning models first predict high-level clinical concepts to inform the final classification, though they require additional annotations and may inadvertently encode unintended information [26]. Case-based models use class-discriminative prototypes for comparison with input features but can be sensitive to noise and difficult to train. Counterfactual explanations perturb input images to generate opposite predictions, yet may produce unrealistic or low-resolution results [27]. Concept attribution methods quantify the influence of high-level image concepts or radiomics features, though annotating such concepts can be challenging and the features may lack reproducibility [25]. Language-based methods provide textual justifications for predictions, which require structured diagnostic reports and extra annotation effort [28]. Latent space interpretation uncovers salient factors in the data relative to clinical knowledge but projecting high-dimensional features into two dimensions may lose information, and similarity in latent space does not always match human-interpretable similarity [29]. Attribution maps highlight regions considered important by the model, though they do not indicate how these regions contribute to the prediction and may overlap across classes [30]. Incorporating anatomical priors introduces task-specific structural knowledge, but requires specialized expertise and cannot be applied universally [31]. Finally, visualizing internal network representations shows the features learned by deep learning blocks, though these patterns can be difficult to interpret in medical images. Collectively, these methods provide diverse avenues for improving model transparency, each with its own strengths and limitations [32].

In summary, the main contributions of this work are as follows. We propose a novel signature construction framework based on deep learning–driven tumor segmentation models, where the segmentation masks and explainability method Grad-CAM are jointly utilized to generate salient maps for defining biologically meaningful signature regions. These signature regions are further used to construct a new type of imaging signature, which is validated through downstream deep learning–based classification models to demonstrate its discriminative effectiveness. To further assess the robustness and interpretability of the proposed signature, radiomic features are extracted from the signature regions and systematically analyzed using statistical machine learning methods and Shapley value–based feature attribution. In addition, the most predictive radiomic features are identified and linked with pathological characteristics to provide biological interpretation of the proposed signature. Finally, extensive experiments are conducted on multiple datasets, including the BUSI breast ultrasound dataset for benign–malignant classification, the BraTS brain tumor dataset for subtype classification, and the KiTS and UF Health renal tumor datasets for benign–malignant classification, to comprehensively validate the effectiveness and generalizability of the proposed framework.

# 2 Results

## 2.1 Signature Validation with Deep Learning

Table. 1 illustrates the classification performance of a deep learning model incorporating radiomics signatures. Based on a comprehensive analysis of deep learning validation, the performance of the four deep learning models—EfficientNet [33], ResNet50 [34], DenseNet121 [35], and Vision Transformer(ViT) [36] —demonstrates a consistent trend where increasing the amount of guided information leads to superior classification across all three datasets. We aim to evaluate the reliability and generalizability of the proposed signature framework across three medical imaging modalities and three tumor classifications: brain tumor subtype on MRI, breast tumor malignancy on ultrasound, and renal tumor malignancy on CT. The experiments are designed with three input settings: image-only, image + segmentation mask, and image + segmentation mask + signature. The image-only setting evaluates the model's intrinsic discriminative ability from raw imaging data. Adding the segmentation mask explicitly highlights the tumor region and guides the model toward lesion-focused prediction. Building on this, incorporating the signature allows us to assess whether the signature provides complementary information beyond the annotated tumor region. Specifically, the signature may capture tumor-associated regions that are not included in the manual annotation, while also excluding regions within the tumor mask that are less relevant to the prediction task. Overall, incorporating signature consistently improved classification performance. Compared with using the image alone, introducing the tumor mask generally increased AUC, precision, and F1 score across most datasets and models, indicating that explicitly highlighting the lesion region enables the network to focus on diagnostically relevant structures. Building upon this, further incorporating the proposed signature produced additional performance gains in the majority of experiments, demonstrating that the signature provides complementary information beyond conventional tumor segmentation.

Across nearly all experiments, incorporating the tumor segmentation mask improved classification performance over using the image alone. On the BraTS dataset, introducing the mask consistently increased AUC for all four backbones, with improvements ranging from 0.2% (DenseNet) to 5.5% (ResNet). Similar gains were observed in Precision and F1 score, indicating that explicitly highlighting the tumor region enables the models to better focus on lesion-related information while suppressing irrelevant background structures. A similar trend was observed on the BUSI dataset, where Image + Mask consistently outperformed Image across all models. Notably, ViT exhibited the largest improvement, with AUC increasing from 0.609 to 0.851 and F1 improving from 0.437 to 0.669, suggesting that transformer-based models benefit substantially from explicit anatomical guidance in ultrasound images. On the UF Health dataset, although the classification task remained challenging, the mask still improved AUC for three of the four backbone networks, particularly for ResNet (0.502 to 0.594), demonstrating that lesion localization provides useful prior information even in complex real-world CT data.

Building upon the segmentation mask, incorporating the proposed signature further improved classification performance in most experiments. On the BraTS dataset, Image + Mask + Signature consistently achieved the highest AUC across all backbone networks, increasing AUC on the setting of Image + Mask from 0.747 to 0.773 (EfficientNet), 0.914 to 0.916 (ResNet), 0.951 to 0.960 (DenseNet), and 0.850 to 0.857 (ViT). Similar improvements were observed in Precision and F1 score, with DenseNet achieving the best overall performance (AUC = 0.960, F1 = 0.866). These results suggest that the proposed signature captures additional discriminative regions beyond the segmented tumor. The benefit of the signature was even more evident on the BUSI dataset. Compared with Image + Mask, incorporating the signature consistently improved all three evaluation metrics across the four backbone networks. For example, DenseNet achieved an AUC improvement from 0.856 to 0.883, while ViT further increased from 0.851 to 0.857, accompanied by substantial gains in Precision (0.697 to 0.790) and F1 (0.669 to 0.717). These findings indicate that the proposed signature effectively complements the tumor mask. On the UF Health renal CT dataset, the improvements introduced by the signature were comparatively smaller but remained generally positive. EfficientNet exhibited the largest gain, with AUC increasing from 0.491 to 0.567, while ViT improved from 0.546 to 0.588. Although ResNet achieved a slightly higher AUC using Image + Mask alone (0.594) than Image + Mask + Signature (0.590), the latter yielded improved Precision and F1 score, suggesting a better balance between sensitivity and specificity. DenseNet showed only marginal changes after incorporating the signature, indicating that renal tumor classification relies more heavily on global tumor morphology and surrounding anatomical context than the other two datasets.

Overall, the results reveal a clear progressive improvement from Image to Image + Mask and finally to Image + Mask + Signature. The segmentation mask consistently guides the networks toward lesion-focused representations, whereas the proposed signature further refines the tumor region by emphasizing diagnostically relevant areas and suppressing less informative regions. Consequently, Image + Mask + Signature achieved the best overall performance in 11 of the 12 model–dataset combinations, demonstrating that the proposed signature provides complementary information beyond conventional tumor segmentation and generalizes well across different imaging modalities, lesion types, and deep learning architectures.

## 2.2 Signature Validation with Machine Learning

Fig. 1 illustrates the classification performance of a deep learning model incorporating radiomics signatures. According to the detailed experimental results in machine learning classification on radiomics features. The performance of the four machine learning models—RandomForest, XGBoost, LinearSVM, and Logistic Regression (LR)—is highly dependent on the specific dataset and the regional strategy used for radiomics feature extraction. The intersection of the signature and the mask is used to identify regions that are jointly highlighted as important by both representations. This setting evaluates whether the signature can refine the tumor mask by excluding non-critical or task-irrelevant regions while preserving the most discriminative tumor areas. The union of the signature and the mask incorporates both the annotated tumor region and the signature-derived regions, allowing us to evaluate whether the signature provides complementary discriminative information beyond the segmentation mask. For the brain tumor MRI dataset (top row), incorporating the proposed signature consistently improved classification performance across all classifiers. The Mask ∩ Signature strategy achieved the highest AUC in every classifier, reaching 0.924, 0.939, 0.944, and 0.947 for Random Forest, XGBoost, LinearSVM, and Logistic Regression, respectively. Compared with using the tumor mask alone, the improvements ranged from 0.6% to 3.0%. Although the signature-only input also produced competitive performance, combining the signature with the tumor mask consistently yielded the best results, suggesting that the signature effectively removes task-irrelevant tumor regions while preserving the most discriminative information. For the BUSI dataset (middle row), a similar trend was observed. The Mask ∩ Signature strategy again achieved the highest AUC for three of the four classifiers, obtaining 0.860, 0.898, and 0.866 for Random Forest, XGBoost, and Logistic Regression, corresponding to improvements of 3.1%, 10.9%, and 3.7% over the tumor mask alone. For LinearSVM, the signature-only setting achieved the highest AUC (0.838), while Mask ∩ Signature produced comparable performance (0.827). These results indicate that the proposed signature provides complementary information beyond the manually annotated tumor region and substantially enhances discriminative capability, particularly for the more heterogeneous ultrasound images. In contrast, for the UF health dataset (bottom row), the benefit of the signature was less consistent. The Mask ∪ Signature strategy produced the highest AUC for Random Forest (0.647), LinearSVM (0.705), and Logistic Regression (0.730), whereas the tumor mask alone remained the best-performing input for XGBoost (0.627). Notably, both the Mask ∩ Signature and signature-only settings generally reduced performance, with the signature-only input exhibiting the largest degradation across all classifiers. This suggests that, unlike MRI and ultrasound, renal tumor classification relies more heavily on complete tumor morphology, and removing portions of the annotated lesion may discard useful diagnostic information. Nevertheless, augmenting the tumor mask with the signature still improved performance for most classifiers, indicating that the signature can provide complementary contextual cues without replacing the original lesion annotation.

Overall, these experiments demonstrate that the proposed signature is highly generalizable across different imaging modalities and lesion types. For MRI brain tumors and ultrasound breast tumors, the signature-guided region consistently provides more discriminative information than the complete tumor mask, indicating that it successfully suppresses irrelevant regions while emphasizing diagnostically important areas. For renal CT, where global tumor morphology appears more informative, preserving both the original tumor mask and the signature yields the best performance. These findings support the hypothesis that the proposed signature captures clinically meaningful information beyond conventional lesion segmentation and can serve as an effective complementary imaging representation across diverse medical imaging tasks.

# 3 Discussion

## 3.1 Ablation Experiment

Fig. 4 illustrates two sets of experiments on the BUSI dataset using ResNet50. The first experiment evaluates the fusion of segmentation masks and saliency maps for tumor classification, while the second compares the performance of single-model configurations against the proposed signature-based fusion strategy. The ablation results demonstrate that both the fusion strategy and the mask source substantially influence downstream classification performance. Among all fusion strategies, the union min configuration achieved the best overall balance, obtaining the highest AUC (0.904), the second-best precision (0.862), and the best F1 score (0.844), indicating that combining the union mask with element-wise minimum saliency fusion can better preserve diagnostically relevant regions while suppressing noisy activations. In contrast, the union_mean and intersection_mean strategies showed noticeably lower precision and F1 values, suggesting that averaging saliency responses may dilute discriminative lesion characteristics and introduce ambiguous feature representations. Interestingly, although inter max and inter mean maintained relatively competitive AUC values, their lower F1 scores indicate reduced robustness in balanced classification performance, implying that intersection-based masks may remove potentially useful contextual regions required for accurate decision-making.

The comparison among different mask types further highlights the importance of region selection quality. Using signature alone consistently achieved the best performance across all metrics, including Accuracy (0.842), AUC (0.878), Precision (0.842), and F1 score (0.796), demonstrating that CAM-based localization provides highly discriminative semantic information for classification. Meanwhile, segmentation masks generated by UNet and UNETR yielded comparatively lower performance, especially in F1 score, indicating that purely anatomical segmentation regions may not fully align with the most classification-relevant features. Nevertheless, the relatively close AUC values among the three mask types suggest that segmentation masks still provide useful structural priors. Overall, these findings suggest that integrating segmentation-derived spatial constraints with discriminative saliency information is beneficial, while overly aggressive averaging or restrictive region intersection may weaken the representation of clinically important patterns.

## 3.2 Interpretable Analysis

Fig. 2 illustrates the top features derived from machine learning models (Random Forest and Logistic Regression) for tumor classification based on radiomics features extracted from the original mask and from the intersection between the mask and the signature. We focus on features that appear in the top-10 ranked feature sets after signature-based processing compared with mask-only features. In particular, we analyze features that are consistently ranked among the top-10 in models' groups (e.g., RF and XGBoost, SVM and LR), while being absent from the top-10 feature list obtained from mask-only radiomics, highlighting the discriminative contribution introduced by the signature-guided region refinement.

**BraTS**: Compared with the original mask (bottom row), the Mask Removing areas identified by the mask but not captured by the CAM (top row) produced a substantial shift in feature importance patterns across LinearSVM, Logistic Regression, Random Forest, and XGBoost, indicating that excluding non-signature regions fundamentally changed the biological basis of model decision-making. Across the four MRI modalities (T1c, T1n, T2w, and T2f), the signature-guided regions consistently introduced radiomic features associated with biologically meaningful tumor subregions that were not fully represented by the original segmentation masks. Rather than emphasizing only tumor bulk characteristics, the enhanced signatures preferentially captured infiltrative margins, heterogeneous vascularized components, edema-associated transition zones, and structurally organized versus disrupted tissue architectures, thereby improving the distinction between glioma (GLI) and meningioma (MEN). Linear machine learning models (LinearSVM and Logistic Regression) primarily emphasized texture-related features, including GLCM DifferenceVariance, suggesting that linear models were more sensitive to local intensity variation and texture complexity patterns across T1c modalities. T1c imaging is primarily used to evaluate contrast-enhancement behavior and vascular heterogeneity. Therefore, intensity distribution–based features extracted from T1c images would reflect the degree of enhancement heterogeneity within the tumor.

- **GLCM_DifferenceVariance** is a GLCM-based texture feature that quantifies the variance of intensity differences between neighboring voxels. It reflects both the magnitude of local intensity fluctuations and the visual heterogeneity of contrast enhancement. Higher values indicate stronger local intensity variations, appearing as irregular, patchy, and heterogeneous enhancement patterns with scattered bright and dark regions, often associated with structural irregularity and complex tissue composition. In contrast, lower values correspond to more uniform and smoothly distributed enhancement with relatively homogeneous intensity appearance.

These findings are consistent with the pathological characteristics of gliomas. Glcm DifferenceVariance reflects infiltrative growth patterns and structural irregularity in Glioma; in contrast, Meningioma tumors typically exhibit more homogeneous, well-circumscribed, and uniform enhancement.

**BUSI**: Tree-based machine learning models rely more on structural and spatial statistical features such as Shape_SurfaceArea, GLCM_Correlation, and NGTDM_ Coarseness, whereas linear machine learning models tend to emphasize finer-grained texture and intensity distribution features, such as GLSZM_SmallAreaLowGrayLevelEmphasis, GLCM_MaximumProbability.

- **Shape_SurfaceArea** represents the total boundary area of the tumor in threedimensional space. A larger surface area corresponds to either a bigger lesion or a more irregular, lobulated outer contour, appearing on ultrasound as a lesion with complex, unsmooth margins rather than a simple round boundary.
- **GLCM_correlation** derived from the gray-level co-occurrence matrix (GLCM), it measures the linear dependency of gray levels between neighboring pixels, defined as the normalized covariance of the joint intensity distribution. Higher values indicate that the intensity of one pixel can be strongly and linearly predicted from its neighbor; visually, this appears on ultrasound as a structured region where brightness transitions smoothly and directionally across the lesion, resembling parallel bands or gradual intensity gradients rather than random scattered bright and dark spots.
- **NGTDM_Coarseness** is derived from the neighborhood gray-tone difference matrix (NGTDM), it quantifies the average absolute difference between each pixel and the mean intensity of its surrounding neighborhood. Higher values indicate minimal deviation between individual pixels and their local neighborhood averages; visually, this appears on ultrasound as broad, homogeneous echo patches where large areas share a similar gray tone, resembling a coarse-grained flat surface, while lower values appear as densely speckled, salt-and-pepper texture where brightness fluctuates rapidly at a fine scale.
- **GLSZM_SmallAreaLowGrayLevelEmphasis** derived from the gray level size zone matrix (GLSZM), it quantifies the prevalence of small-sized regions with low gray-level intensities. Higher values appear on ultrasound as scattered, punctate hypoechoic foci distributed throughout the lesion rather than large confluent hypoechoic zones.
- **GLCM_MaximumProbability** reflects the dominance of the most frequently occurring gray-level pair within the texture. In ultrasound images, a higher Maximum Probability reflects the dominance of the most frequently occurring gray-level pair within the texture. A higher GLCM_MaximumProbability value appears on ultrasound as a visually monotonous region where a single echo intensity predominates, giving the lesion a flat, uniform appearance with minimal internal echo variation.

In general, higher Shape_SurfaceArea is generally associated with malignant lesions, reflecting increased lesion extent and more irregular, infiltrative margins commonly observed in breast ultrasound. Higher NGTDM_Coarseness are more commonly associated with benign lesions, reflecting more homogeneous and structurally regular echotexture, whereas malignant lesions tend to show reduced values due to increased heterogeneity and fine-grained textural complexity. And higher GLSZM_SmallAreaLowGrayLevelEmphasis suggests the presence of small low-intensity (hypoechoic or anechoic) regions, which may correspond to cystic degeneration, necrotic-like changes, or other forms of internal heterogeneity commonly observed in malignant lesions. A higher Maximum Probability in ultrasound images indicates that a specific pair of co-occurring intensity values dominates the GLCM distribution, reflecting the predominance of homogeneous low-echo regions commonly observed in malignant lesions.

**UF Health**: Tree-based machine learning models primarily rely on features such as GLSZM_LargeAreaLowGrayLevelEmphasis, Shape_SurfaceVolumeRatio, and Shape _Maximum3DDiameter, which capture global morphological characteristics, lesion extent, and the presence of large low-intensity heterogeneous regions. In contrast, linear machine learning models tend to utilize features including Firstorder_10Percentile, Shape_Elongation, Shape_Maximum2DDiameterColumn, and GLSZM_LargeAreaEmphasis, which mainly reflect intensity distribution statistics, directional shape properties, and more localized or single-scale structural patterns.

- **GLSZM_LargeAreaLowGrayLevelEmphasis** reflects large contiguous regions with uniformly low Hounsfield Unit values, appearing as extensive dark, hypoattenuating zones on CT images.
- **Firstorder_interquartileRange** is a robust statistical measure of dispersion defined as the difference between the 75th and 25th percentiles of the Hounsfield Unit distribution, representing the spread of the central 50% of HU values within the lesion. A larger IQR corresponds to a wider range of attenuation values, appearing as a heterogeneous mixture of bright and dark regions on CT.
- **Firstorder_10Percentile** represents the 10th percentile of the Hounsfield Unit distribution within the lesion, capturing the lower-end attenuation values. Higher values indicate that even the darkest regions of the lesion maintain relatively elevated HU, appearing as uniformly intermediate-to-high density on CT.
- **Shape_Elongation** describes how elongated the lesion is in three-dimensional space. Higher values correspond to a stretched, asymmetric morphology with a notably larger dimension along one axis, appearing as an elongated or irregular shape rather than a round or oval lesion on CT.

In CTs, GLSZM_LargeAreaLowGrayLevelEmphasis is higher in malignant lesions, indicating more extensive low-intensity heterogeneous regions reflecting necrosis, cystic change, liquefaction, or low-cellularity areas; and Shape_Elongation is higher in malignant lesions, suggesting more directional and infiltrative growth patterns. Lower Firstorder_10Percentile values indicate a greater proportion of hypoattenuating components, which may be associated with necrosis, fluid content, or less dense tumor tissue. A larger interquartile range reflects greater tissue heterogeneity, suggesting mixed compositions such as fat, soft tissue, fluid, or calcification coexisting within the lesion, which is more commonly associated with benign characteristics.

### 3.3 Limitation

Despite the effectiveness of the proposed multimodal fusion framework in integrating CT images, segmentation masks, and Grad-CAM–based attention maps for 3D tumor classification, several limitations should be acknowledged. First, the framework is highly dependent on the accuracy of the underlying segmentation models (nnU-Net and UNETR), and any segmentation errors may propagate through CAM generation and radiomic feature extraction, ultimately affecting classification performance. Second, although the mutual information–based adaptive thresholding strategy enables case-specific refinement of attention regions, it may still inadvertently exclude subtle but clinically relevant tumor areas or retain irrelevant background signals, thereby introducing variability in the extracted signatures. Third, because the framework requires one subset of the data for training segmentation models and another subset for downstream classification and prediction, the effective sample size available for each stage is relatively limited. This constrained data availability may reduce the full potential of the framework in discovering and mining robust imaging signatures, particularly for complex or heterogeneous lesions. Larger-scale datasets and multicenter cohorts may further enhance the stability and generalizability of the discovered signatures. Fourth, the requirement for complete multimodal inputs (CT, segmentation mask, and signature) may introduce selection bias by excluding incomplete cases, potentially limiting generalizability to real-world clinical scenarios where missing data are common. Fifth, despite the use of class-weighted loss functions and weighted sampling strategies, residual class imbalance may still affect model sensitivity, particularly for underrepresented categories. Sixth, although Grad-CAM provides valuable interpretability by highlighting discriminative regions, it reflects model attention rather than true causal pathology, and therefore highlighted regions should be interpreted with caution in clinical contexts. Finally, the current framework was constructed using only two segmentation models together with Grad-CAM for saliency map generation. Although the results demonstrate the feasibility and effectiveness of the proposed strategy, a broader exploration of segmentation backbones and saliency generation techniques was not conducted. More advanced or diverse segmentation architectures, as well as alternative explainability methods such as Score-CAM, Layer-CAM, EigenCAM, or attention-based visualization approaches, may further improve the quality of the generated saliency representations and potentially enhance downstream predictive performance. Therefore, there remains considerable room for future exploration and optimization of the framework design.

## 4 Method

### 4.1 Datasets

We evaluate our framework on three publicly available medical imaging datasets, including BUSI, BraTS, KiTS and UF Health(private dataset), which cover different imaging modalities and tumor types. The BUSI dataset is a breast ultrasound dataset widely used for tumor classification tasks. It contains both benign and malignant cases, and is commonly used to evaluate binary tumor classification performance under limited data settings. The BraTS dataset is a multi-institutional brain tumor MRI dataset that includes two main cohorts used in this study: glioma (GLI) and meningioma (MEN). The GLI cohort consists of glioma cases with varying grades of aggressiveness, while the MEN cohort includes meningioma cases. These two subgroups represent distinct brain tumor subtypes with different pathological characteristics and are primarily used for tumor segmentation and related downstream analysis. The KiTS dataset consists of abdominal CT scans focused on kidney tumor segmentation. It includes cases with and without visible tumor presence, and is widely used for evaluating tumor segmentation and detection performance in abdominal imaging. UF Health is the academic health system of the University of Florida (UF), one of the largest and most comprehensive university-affiliated healthcare networks in the United States. It integrates clinical care, medical education, and biomedical research within a single system. In this study, we utilize renal cell carcinoma CT data from UF Health for benign–malignant tumor classification analysis.

For the BUSI dataset, we construct the training split from the original dataset while ensuring a balanced class distribution across categories. The segmentation model is trained on this subset to provide consistent tumor region localization for subsequent analysis; For the BraTS dataset, we follow the official training protocol and use the provided training set to train the segmentation model. The official validation set is used for classification evaluation. To enable unified tumor representation, we define a generalized tumor label by merging all intra-tumoral subregions into a single class, following a similar design philosophy to the MEN cohort setting, thereby obtaining a pan-tumor representation for downstream analysis; For renal tumor analysis, we focus on the KiTS dataset for training the segmentation model, while using external UF Health renal cell carcinoma CT data for classification validation. Due to the presence of significant anatomical noise in abdominal CT images, including surrounding organs and potential metastatic regions, we adopt a official KiTS-pretrained nnU-Net model to perform segmentation on UF Health data. Only the regions corresponding to kidney and tumor are retained for subsequent classification and signature extraction, thereby reducing interference from irrelevant anatomical structures. During segmentation model training, we retain only the tumor label and exclude other classes such as kidney parenchyma and cysts. This design ensures that the learned segmentation model focuses exclusively on tumor-specific information, enabling the generated signatures to capture tumor characteristics without contamination from surrounding non-tumor tissues.

## 4.2 Signature Construction

Building upon the nnU-Net framework, we employ two complementary segmentation models, nnU-Net [37] and UNETR [38], to generate robust representations of tumor regions. For each model, Gradient-weighted Class Activation Mapping (GradCAM) [39] is applied only to the final decoder layer to obtain the saliency map. To further enhance robustness, we perform cross-model fusion at both the segmentation and saliency levels. For segmentation masks, we compute the union between nnUNet and UNETR predictions, representing conservative and inclusive tumor regions, respectively. For saliency maps, three fusion strategies are explored—element-wise minimum to capture varying degrees of agreement and emphasis between models. The Signature is extracted from each CAM by searching for an optimal binarization threshold that maximizes mutual information with the corresponding segmentation Mask. To focus saliency activations on the tumor region, an exponential distance decay is first applied to the CAM based on voxel-wise (3D) or pixel-wise (2D) distances from the Mask centroid, defined as:

$$cam_decayed = cam \times exp(-dist \times \lambda) \qquad (1)$$

where the decay coefficient $\lambda = \frac{1}{r/D}$ is adaptively determined by the equivalent radius of the Mask and the maximum image dimension $D$. For 3D volumes, the equivalent radius is derived from the spherical volume formula $r = \left(\frac{3V}{4\pi}\right)^{1/3}$, where $V$ is the Mask voxel count; for 2D images, it is derived from the circular area formula $\lambda = \frac{1}{r/D}$, where $A$ is the Mask pixel count. In both cases, smaller lesions yield a larger $\lambda$, producing steeper decay that more effectively suppresses distal spurious activations. The decayed CAM is quantized to uint8, and a histogram-based search evaluates all 256 candidate thresholds by constructing a $2 \times 2$ joint probability matrix at each threshold and computing mutual information as:

$$I = \sum p(x,y) \log \frac{p(x,y)}{p(x)p(y)} \qquad (2)$$

with cumulative histogram summation reducing the search complexity from $O(N \times 256)$ to $O(256)$ for efficient large-scale parallel processing. The threshold yielding maximum mutual information is used to binarize the CAM, and the resulting region — representing the saliency activation with the highest informational overlap with the Mask — is retained as the final Signature.

To refine attention localization, a mutual information–based adaptive thresholding strategy is introduced. Specifically, within the region of interest defined by the segmentation mask, candidate saliency maps' thresholds (0–255) are exhaustively evaluated. For each threshold, the binarized CAM is compared with the ground-truth mask using mutual information computed from their joint histogram. The threshold that maximizes mutual information is selected as the optimal case-specific cutoff, enabling data-driven identification of discriminative tumor regions for feature extraction.

The resulting thresholded saliency map is binarized such that values above the threshold are set to 1 and those below are set to 0, yielding signature-defined tumor regions used for downstream classification and interpretability analysis.

## 4.3 Tumor Prediction

We propose a multimodal fusion framework for three-dimensional (3D) tumor classification, aiming to distinguish benign and malignant cases. For each subject, three modalities are collected from a standardized directory: original images, segmentation masks, and signatures generated from trained models. Only samples with complete modalities are retained, and data are organized into benign and malignant classes (labels 0 and 1). To ensure reproducibility, random seeds are fixed across Python, NumPy, PyTorch, MONAI, and data-loading workers, enabling fully deterministic training.

A unified 3D preprocessing pipeline is implemented using MONAI. NIfTI data are converted to channel-first tensors, and intensity normalization is applied to CT images. All modalities are resampled to a fixed resolution using trilinear interpolation for CT and nearest-neighbor interpolation for masks and signatures. For the BUSI dataset, the input images are of size 512×512, while the BraTS dataset consists of 3D volumes with spatial dimensions of 256×256×256, and the KiTS dataset contains 3D volumes of size 128 × 128 × 128. We design a progressive multimodal fusion strategy with three input settings: (1) CT only, (2) CT + segmentation mask, and (3) CT + mask + Signature. This design allows progressive integration of anatomical structure and attention priors to enhance discriminative learning for tumor classification.

Multiple backbone architectures are evaluated, including CNN-based models (DenseNet121, ResNet50, EfficientNet) and a Transformer-based model (ViT), all adapted for 3-channel input and binary classification. To address class imbalance, we combine class-weighted cross-entropy loss with weighted random sampling. Additionally, Focal Loss [40] (gamma = 3) is adopted to emphasize hard-to-classify samples and improve robustness. For evaluation, stratified 8:2 train–test splits are used, and experiments are repeated over multiple independent runs with different random seeds. Performance is assessed using accuracy, macro-F1, macro-precision, and AUC. Final results are reported as mean ± standard deviation to quantify stability and generalization. Overall, the proposed framework integrates segmentation priors and signature-derived attention into a unified classification pipeline, improving both predictive performance and interpretability for tumor classification tasks.

### 4.4 Interpretability

Radiomic features are extracted from tumor regions defined by the binarized signatures and segmentation mask for classification using PyRadiomics [41]. These features are used to train a Random Forest, XGBoost, Support vector networks (SVM), and Logistic Regression, and performance is evaluated over 5 independent 80/20 train-test splits. The final performance is reported as mean ± standard deviation of AUC. To interpret model behavior, SHAP analysis is performed to quantify feature contributions to tumor classification. The SHAP summary plot reveals that predictive signals are primarily driven by three categories of radiomic features: first-order intensity statistics, texture heterogeneity, and structural pattern descriptors.

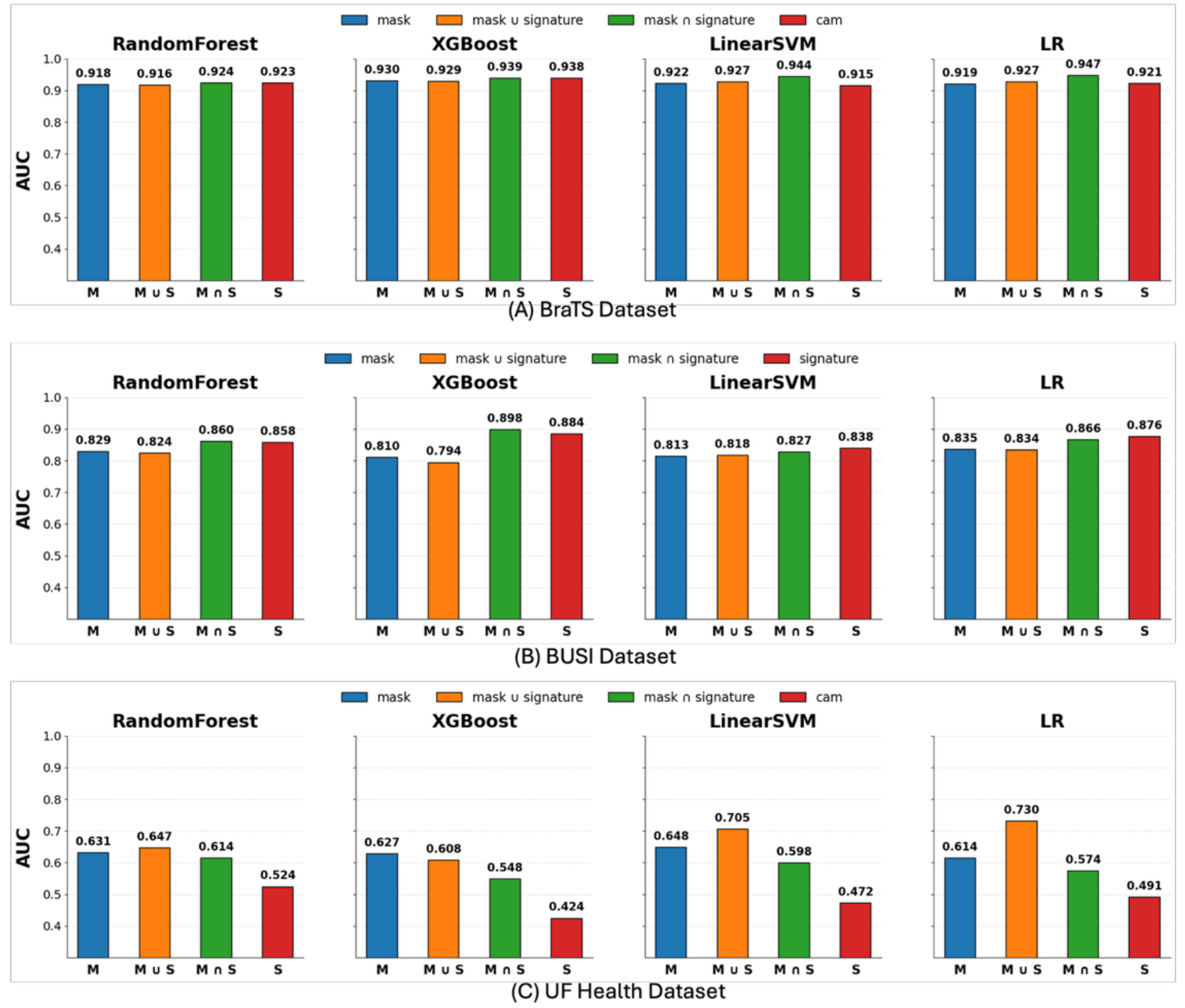

**Fig. 1** Comparison of classification performance using different signature-guided region definitions. Classification performance (AUC) of four machine learning classifiers (Random Forest, XGBoost, LinearSVM, and Logistic Regression (LR)) under four input settings: mask (M), mask ∪ signature (M ∪ S), mask ∩ signature (M ∩ S), and signature only (S). The three rows correspond to the BraTS, BUSI, and UF Health datasets, respectively. The mask ∪ signature setting evaluates whether the signature provides complementary discriminative information beyond the annotated tumor region, whereas the mask ∩ signature setting assesses whether the signature refines the tumor region by removing task-irrelevant areas while preserving diagnostically important information.

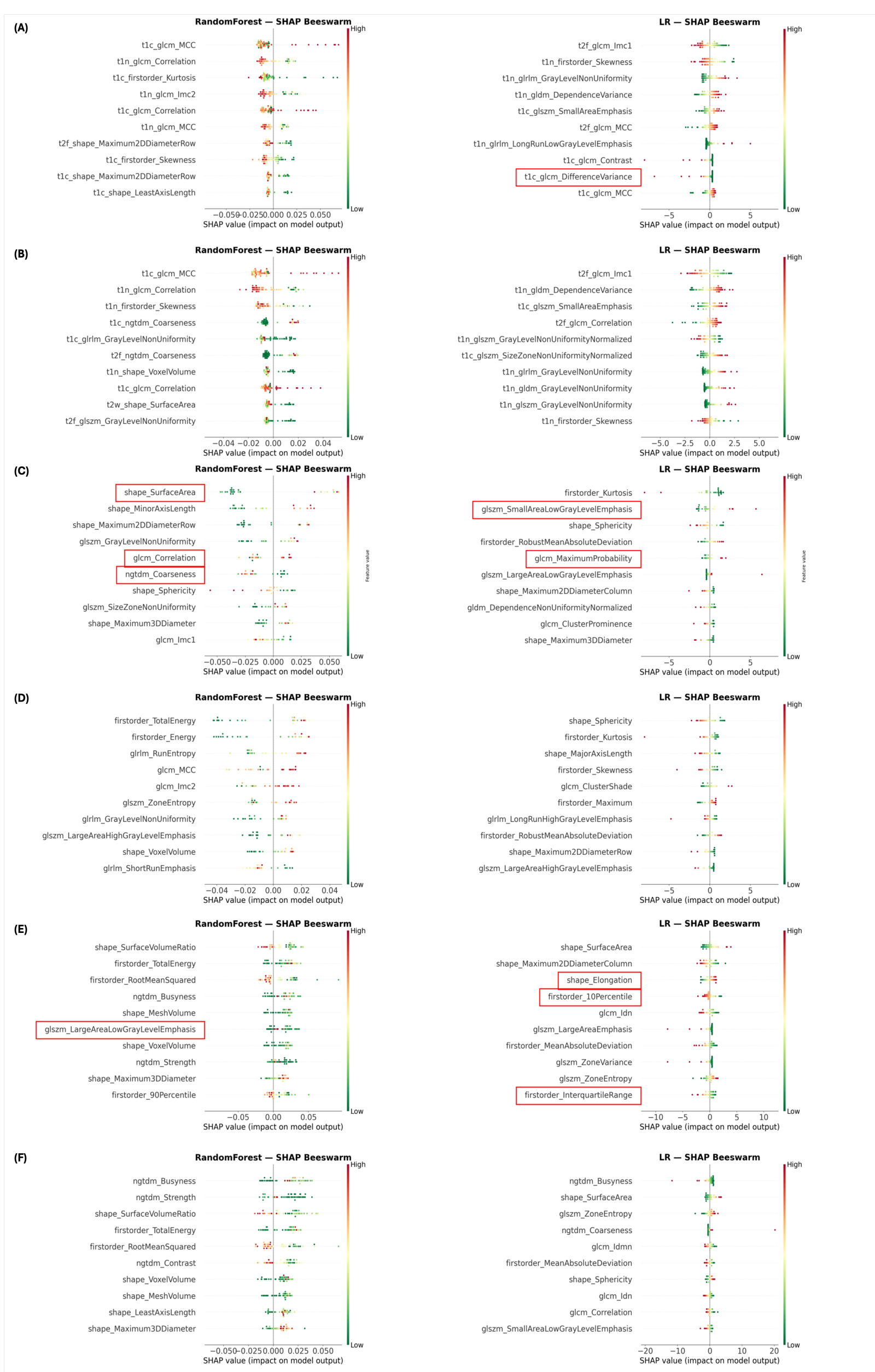

**Fig. 2** SHAP analysis of radiomic features before and after signature refinement. SHAP beeswarm plots of the top 10 radiomic features ranked by SHAP importance using Random Forest (left) and Logistic Regression (LR) (right). (A–B) Results obtained from the BraTS dataset, (C–D) results from the BUSI dataset under and (E-F) results from the UF Health dataset. Red boxes highlight representative radiomic features whose importance changes after incorporating the proposed signature, indicating that the signature modifies the contribution of both texture- and shape-related features. (B), (D), and (F) show radiomic features extracted from the tumor mask, while (A) and (C) show radiomic features extracted from the intersection of the mask and the signature, and (E) shows radiomic features extracted from the union of the mask and the signature.

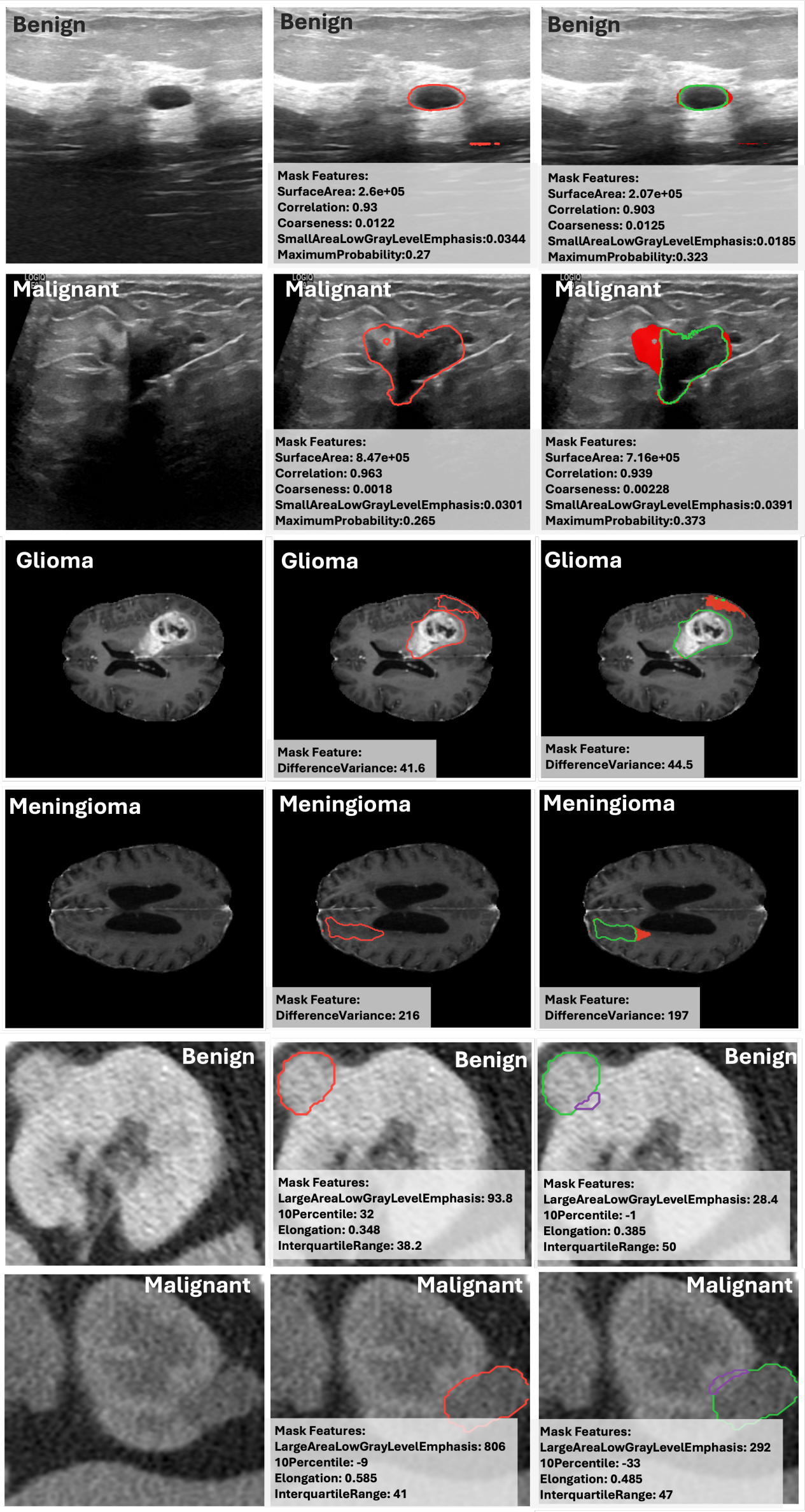

**Fig. 3** Visualization of top radiomics features and region-level overlap between segmentation masks and signatures across different datasets. The first column displays the corresponding top features within text boxes. The second column shows the lesion region delineated by the red contour of the segmentation mask. In the third column, different visualization strategies are applied across datasets to illustrate the relationship between mask and signature regions. For the BUSI ultrasound dataset (first and second rows), the green region represents the intersection between the mask and the signature, while the red region highlights the removed mask area outside the signature. For the BraTS brain tumor dataset (third and fourth rows), a similar mask–signature interaction is visualized to reflect tumor-related regions in MRI scans. For the kidney tumor dataset (fifth and sixth rows), the purple contour indicates regions where the union of mask and signature exceeds the original mask, highlighting additional activated areas beyond the segmented lesion.

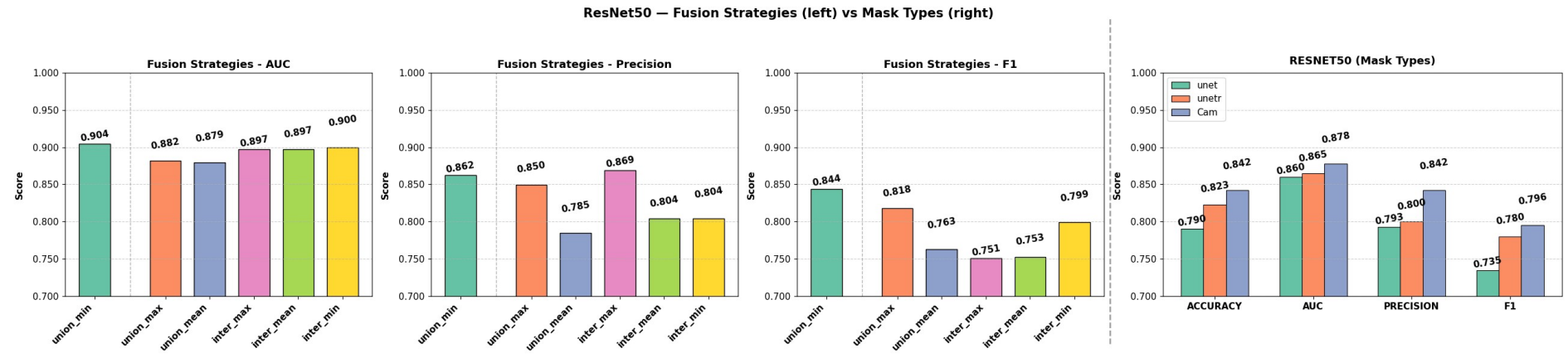

**Fig. 4** Ablation study based on BUSI dataset and ResNet50 for tumor classification. The first three tables report the performance of six different combinations, where segmentation masks from two models are combined and fused with saliency maps to generate final representations. Specifically, the intersection of masks is paired with saliency maps using three fusion strategies (minimum, mean, and maximum), and the union of masks is further combined with the fused saliency maps, resulting in six experimental settings. The evaluation metrics include AUC, precision, and F1-score. The fourth figure compares the performance of using only nnUNet and only UNETR against the fused signature representation. In this setting, all models are trained using a unified input consisting of image, mask, and signature fusion.

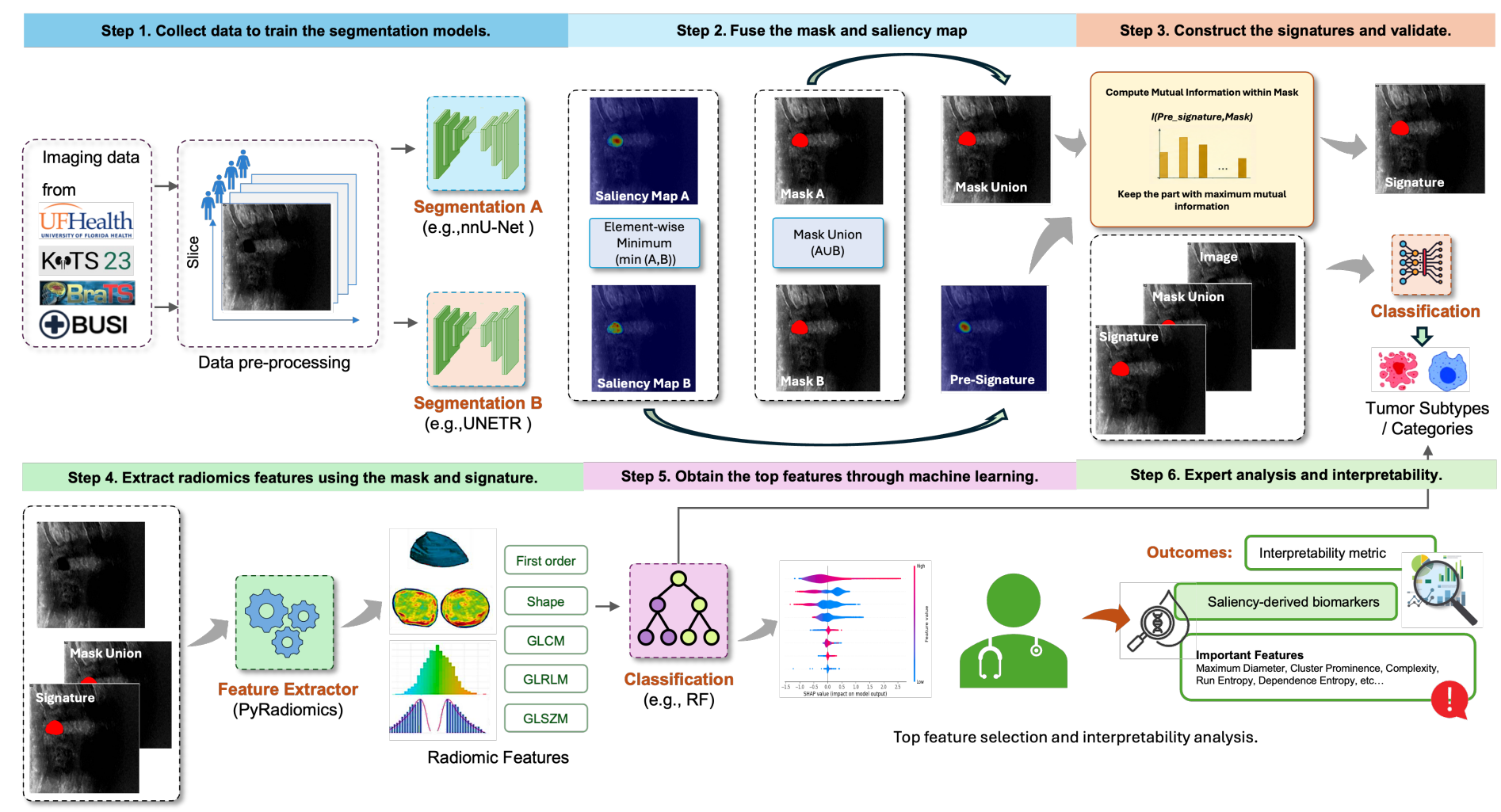

**Fig. 5** The pipeline consists of six steps. In Step 1, medical images are collected from multiple datasets (UF Health, KiTS23, and BraTS) to train two independent segmentation models (Segmentation A and Segmentation B). In Step 2, the saliency maps generated from each segmentation model are fused via element-wise minimum operation, and the corresponding binary masks are combined through union to produce the Mask Union and Pre-Signature regions. In Step 3, mutual information between the Pre-Signature and the Mask Union is computed across candidate regions, and the region with maximum mutual information is retained as the final Signature; the Image, Mask Union, and Signature are then fed into a classifier for tumor subtype or malignancy classification. In Step 4, radiomics features are extracted from both the Mask Union and the Signature regions using a feature extractor, yielding first-order, shape, GLCM, GLRLM, and GLSZM feature sets. In Step 5, machine learning models are applied to select the most discriminative features, with feature importance visualized via SHAP values. In Step 6, an expert interprets the outcomes, where the top-ranked features serve as saliency-derived biomarkers to provide clinically meaningful and interpretable results.

| Dataset | Model | Fusion | AUC | Precision | F1 |
|---|---|---|---|---|---|
| BraTS | EfficientNet | Image | 0.726±0.156 | 0.699±0.122 | 0.637±0.143 |
| | | Image + M | 0.747±0.166 | 0.688±0.118 | 0.639±0.133 |
| | | Image + M+ S | **0.773±0.115** | **0.715±0.103** | **0.650±0.151** |
| | ResNet | Image | 0.866 ± 0.055 | 0.770 ± 0.136 | 0.702 ± 0.167 |
| | | Image + M | 0.914 ± 0.050 | 0.818 ± 0.049 | 0.774 ± 0.103 |
| | | Image + M+ S | **0.916±0.056** | **0.847±0.069** | **0.788± 0.156** |
| | DenseNet | Image | 0.949 ± 0.026 | 0.857 ± 0.055 | 0.823 ± 0.104 |
| | | Image + M | 0.951 ± 0.035 | 0.872 ± 0.060 | 0.832 ± 0.113 |
| | | Image + M+ S | **0.960±0.023** | **0.875±0.046** | **0.866± 0.047** |
| | ViT | Image | 0.831 ± 0.064 | 0.747 ± 0.061 | 0.730 ± 0.061 |
| | | Image + M | 0.850 ± 0.060 | 0.775 ± 0.054 | **0.765± 0.052** |
| | | Image + M+ S | **0.857±0.061** | **0.786±0.064** | 0.762 ± 0.090 |
| BUSI | EfficientNet | Image | 0.841 ± 0.073 | 0.793 ± 0.076 | 0.681 ± 0.075 |
| | | Image + M | 0.849 ± 0.080 | 0.828 ± 0.077 | 0.768 ± 0.073 |
| | | Image + M + S | **0.856±0.078** | **0.853±0.074** | **0.803± 0.065** |
| | ResNet | Image | 0.824 ± 0.081 | 0.725 ± 0.119 | 0.697 ± 0.110 |
| | | Image + M | 0.857 ± 0.069 | 0.825 ± 0.075 | 0.782 ± 0.084 |
| | | Image + M+ S | **0.878±0.069** | **0.842±0.066** | **0.796± 0.096** |
| | DenseNet | Image | 0.819 ± 0.069 | 0.773 ± 0.078 | 0.705 ± 0.087 |
| | | Image + M | 0.856 ± 0.067 | 0.835 ± 0.073 | 0.816 ± 0.070 |
| | | Image + M + S | **0.883±0.061** | **0.845±0.079** | **0.823± 0.068** |
| | ViT | Image | 0.609 ± 0.089 | 0.403 ± 0.110 | 0.437 ± 0.063 |
| | | Image + M | 0.851 ± 0.058 | 0.697 ± 0.176 | 0.669 ± 0.159 |
| | | Image + M+ S | **0.857±0.057** | **0.790±0.090** | **0.717± 0.115** |
| UF Health | EfficientNet | Image | 0.494 ± 0.112 | 0.505 ± 0.057 | 0.434 ± 0.136 |
| | | Image + Mask | 0.491 ± 0.146 | 0.475 ± 0.116 | 0.442 ± 0.121 |
| | | Image + M+ S | **0.567±0.107** | **0.530±0.057** | **0.464± 0.118** |
| | ResNet | Image | 0.502 ± 0.091 | 0.492 ± 0.074 | 0.486 ± 0.050 |
| | | Image + M | **0.594±0.119** | 0.510 ± 0.090 | 0.507 ± 0.070 |
| | | Image + M+ S | 0.590 ± 0.098 | **0.549±0.152** | **0.515± 0.071** |
| | DenseNet | Image | 0.513 ± 0.112 | 0.510 ± 0.080 | **0.506± 0.064** |
| | | Image + M | 0.528 ± 0.076 | 0.497 ± 0.062 | 0.498 ± 0.056 |
| | | Image + M+ S | **0.536±0.117** | **0.512±0.129** | 0.487 ± 0.057 |
| | ViT | Image | 0.542 ± 0.108 | 0.511 ± 0.052 | 0.493 ± 0.067 |
| | | Image + M | 0.546 ± 0.125 | 0.524 ± 0.062 | 0.505 ± 0.080 |
| | | Image + M + S | **0.588±0.100** | **0.527±0.050** | **0.514± 0.061** |

**Table 1.** Classification performance on the BraTS, BUSI, and UF Health datasets using deep learning models. Results are reported as mean ± standard deviation. Note: M is Mask, S is Signature